\documentclass[fleqn,twoside]{article}
\usepackage{amsmath}
\usepackage{amssymb}
\usepackage{espcrc2}

\newcommand{\bracket}[2]{\langle #1|#2\rangle}
\newcommand{\matelem}[3]{\langle #1|#2|#3\rangle}
\newcommand{\measure}[1]{\mathcal{D}#1}
\def \as {\relax\ifmmode\alpha_s\else{$\alpha_s${ }}\fi}

\usepackage{graphicx}
\usepackage[figuresright]{rotating}

\title{On next-to-eikonal exponentiation}

\author{
E. Laenen\address[Nikhef]{Nikhef,  Science Park 105,  1098 XG Amsterdam; \\
  ITF, Utrecht University, Leuvenlaan 4, 3584 CE Utrecht; \\
  ITFA, University of Amsterdam, Science Park 904, 1090 GL
  Amsterdam, The Netherlands},
L. Magnea\address[INFN Torino]{CERN, PH Department, TH Group,
  1211 Geneva 23, Switzerland; \\
  Dipartimento di Fisica Teorica, Universit\`a di Torino,\\
  INFN, Sezione di Torino, Via P. Giuria 1, I-10125 Torino, Italy}, 
G. Stavenga\address[Fermilab]{Fermi National Accelerator Laboratory,   
  MS106, P.O. Box 500, IL 60510, U.S.A.} and
C.D. White\address[IPPP]{Institute for Particle Physics Phenomenology, 
  Department of Physics,\\ Durham University, Durham DH1 3LE, 
  United Kingdom}}
     
\begin{document}

\begin{abstract}
The eikonal approximation is at the heart of many theoretical
and phenomenological studies involving multiple soft gauge boson
emissions in high energy physics. We describe our efforts towards the
extension of  the eikonal approximation for scattering amplitudes to 
the first subleading power in the soft momentum.
\vspace{1pc}
\end{abstract}

\maketitle

\section{Introduction}
\label{sec:introduction}

It is well-known that soft gauge boson emissions often give rise to 
large corrections to hard scattering cross-sections. Generically, if 
$1- x$ is a dimensionless variable related to the energy carried by 
undetected soft gauge bosons in a given process, the differential 
cross-section receives perturbative corrections of the form
\begin{multline}
  \frac{d \sigma}{d x} = \sum_{m, n} \alpha^n \Bigg[
  c_{n m} \left(\frac{\log^m (1- x)}{1- x} \right) \Bigg. \\
  \Bigg. \hspace{-5mm} + \, d_{n m} \log^m (1- x) + 
  \ldots \Bigg] \, ,
\label{soft}
\end{multline}
where $\alpha$ is the coupling constant and generically 
$m \leq 2 n - 1$. The first term contains contributions that are not 
integrable as $x \to 1$, displaying an IR divergence that must be 
cancelled by virtual corrections, and the ellipsis denotes terms
that are suppressed by powers of $1 - x$. When $x \to 1$, the 
convergence of the perturbative expansion breaks down and 
resummation becomes necessary: one would like to know the 
coefficients $\{c_{n m}\}$ and $\{d_{n m}\}$ for all values of $n$. 
Much is known about the $c_{n m}$ coefficients, which are related to the eikonal approximation for soft radiation. This 
amounts to taking in each diagram the leading power term when 
all soft gauge boson momenta $k_i \to 0$.  A full understanding 
of the $d_{n m}$ coefficients requires the use of  the
next-to-eikonal (NE) approximation, in which $k_i \to 0$ for all 
but one gluon, whose momentum is kept to first subleading order 
in the scattering amplitude.

A crucial result for resummations based on the eikonal approximation 
is the fact that amplitudes for soft gauge bosons form an exponential
(``exponentiate''). For abelian gauge theory this has been understood
since the early 1960s~\cite{Yennie:1961ad}. For non-abelian theories,
remarkably, similar results hold \cite{Sterman:1981jc,Gatheral:1983cz,Frenkel:1984pz,Berger:2003zh}.

We have revisited this exponentiation with an eye to extending the 
eikonal approximation, and the results that are built upon it, to 
next-to-leading power in the soft energy/momentum. We begin 
by describing a simple ansatz which generalizes threshold 
resummation, incorporating some recent insights into the structure 
of NE terms in the QCD splitting functions, as well as some 
well-understood phase space effects. While this ansatz is successful 
in reproducing the bulk of NE terms for inclusive cross sections, it
does not give the full answer. We then go on to describe recent more 
systematic attempts to organize all NE terms, using either a path 
integral or a diagrammatic approach.

\section{Extended threshold resummation}
\label{sec:extend-thresh-resumm}

The authors of the three-loop calculation of Altarelli-Parisi
splitting functions~\cite{Moch:2004pa} found a remarkable relation
between eikonal and next-to-eikonal contributions: taking Mellin 
moments of splitting functions, they noted that the coefficients
of terms proportional to $\ln N/N$ are determined by the coefficients
of eikonal logarithms, $\ln N$. Subsequently, Dokshitzer, Marchesini 
and Salam (DMS)~\cite{Dokshitzer:2005bf}, proposed a modified
evolution equation for parton distributions that connects eikonal
and sub-eikonal terms in the splitting function in a nontrivial way,
providing a justification for the results of~\cite{Moch:2004pa}.
In \cite{Laenen:2008ux}, we used these results to extend the 
threshold resummation formulae of \cite{Sterman:1986aj,Catani:1989ne,ELM}, including the modified DMS evolution equation, and
taking into account threshold kinematics at NE level.
For the Mellin moments of the Drell-Yan partonic cross section, 
$\widehat{\omega}(N)$, we proposed the resummed expression
\begin{multline}
\label{eq:1}
  \ln \Big[ \widehat{\omega} (N) \Big] =  
  {\cal F}_{\rm DY} \left( \as (Q^2) \right) + \\
  \int_0^1 \, dz \, z^{N - 1} \, \Bigg\{ \frac{1}{1 - z} \, 
  D \left[ \as \left( \frac{(1 - z)^2 Q^2}{z} \right) \right] \\ + \,
  2 \,\int_{Q^2}^{(1 - z)^2 Q^2/z} \, 
  \frac{d q^2}{q^2} \, P_s \Big[ z, \as (q^2) \Big] \Bigg\}_+ \, ,
\end{multline}
where the $n^{\rm th}$ order term in the DMS-improved space-like 
evolution kernel is given by
\begin{multline}
    \hspace{-1mm}
    P_s^{(n)} (z) = \frac{z}{1 - z} A^{(n)} + C^{(n)} \ln (1 - z) 
    + \ldots \, ,
    \label{modP}
\end{multline}
The function $\mathcal{F}_{\rm DY}$ controls $N$-independent 
terms according to \cite{ELM}, and the function $D$ controls 
contributions from wide-angle soft radiation. The $1/z$ factors in 
the argument of the coupling in the function $D(\alpha_s)$, as
well as the upper limit of the $q^2$ integral, reflect a more accurate
accounting of threshold kinematics, and also lead to $1/N$ effects. 
For DIS we proposed a similar form. 

To assess the quality of our proposal, we compare the expansion
of Eq.~(\ref{eq:1}) in powers of $\alpha_s(Q^2)$, up to two loops, 
with the exact results of \cite{DYexact}, in terms of the coefficients 
$a_{n m}$ and $b_{nm}$ in the expression
\begin{multline}
    \widehat{\omega} (N) = \sum_{i = 0}^\infty \left( \frac{\as}{\pi}
    \right)^n \Bigg[ \sum_{m = 0}^{2 n} \, a_{n m} \ln^m \bar{N} 
    \\ + \, \sum_{m = 0}^{2 n - 1} b_{n m} \frac{\ln^m \bar{N}}{N} 
    \Bigg] + {\cal O} \left( \frac{\ln^p N}{N^2} \right) \, , 
\end{multline}
where $\bar{N} = N\exp(\gamma_E)$. As expected, all $a$
coefficients are reproduced.  At the $1/N$ level we find the 
results shown in table \ref{tab:dycoeff}.
\begin{table*}[hbt]
 \newcommand{\m}{\hphantom{$-$}}
 \newcommand{\cc}[1]{\multicolumn{1}{c}{#1}}
 \renewcommand{\tabcolsep}{1pc} % enlarge column spacing
 \renewcommand{\arraystretch}{1.3} % enlarge line spacing
 \caption{\label{tab:dycoeff} 
Comparison of exact and resummed 2-loop coefficients for the 
Drell-Yan cross section. For each color structure, the left column 
contains the exact results, the right column contains the prediction 
derived from resummation according to Eq.~(\ref{eq:1}).}
\begin{tabular}{|c|c|c|c|c|c|c|}
\hline
    & \multicolumn{2}{c|}{$C_F^2$}  &  \multicolumn{2}{c|}{$C_AC_F$}  
    & \multicolumn{2}{c|}{$n_fC_F$} \\ 
  \hline
  \hline
   $b_{23}$   &   $4$   &   $4$   &   $0$   &   $0$   &   $0$   &   
   $0$   \\
   $b_{22}$   & $ \frac{7}{2}$  & 4 & $\frac{11}{6}$ & 
   $\frac{11}{6}$ & 
   $- \frac{1}{3}$ & $- \frac{1}{3}$  \\
   $b_{21}$   &   $8 \zeta_2 - \frac{43}{4}$   &   
   $8 \zeta_2 - 11$   &   
   $- \zeta_2 + \frac{239}{36} $   &   $- \zeta_2 + \frac{133}{18} 
   $   &   $- \frac{11}{9}$   &   $- \frac{11}{9}$   \\
   $b_{20}$   &   $- \frac{1}{2} \zeta_2 - \frac{3}{4}$   &   
   $4 \zeta_2$   &   
   $- \frac{7}{4} \zeta_3 + \frac{275}{216}$   &   
   $\frac{7}{4} \zeta_3 + \frac{11}{3} \zeta_2 - \frac{101}{54}$   &   
   $- \frac{19}{27}$   &     $- \frac{2}{3} \zeta_2 + \frac{7}{27}$   
   \\ \hline
\end{tabular}
\end{table*}
We see that the leading $1/N$ terms ($b_{23}$) are reproduced 
for each color structure, while an excellent approximation for the 
next-to-leading ones ($b_{22}$) is reached, and even $b_{21}$
is well reproduced. A similar conclusion holds for the DIS case, 
where we could  compare with 2-loop  \cite{Zijlstra:1992qd} and 
even 3-loop results \cite{Vermaseren:2005qc}.

Full agreement at NE accuracy is, however, not reached here, nor
in other approaches \cite{vogt1,soar,mv,grun1,grun2,grun3,grun4}. 
To this end, a deeper understanding of exponentiation at NE accuracy 
is called for.

\section{Path integral approach}
\label{sec:path-integr-appr}

It is possible to cast the exponentiation all possible subdiagrams
involving soft gauge boson exchanges between external charged 
energetic lines in terms of the textbook exponentiation of connected 
Feynman diagrams~\cite{Laenen:2008gt}. This technique in fact encompasses both the eikonal and next-to-eikonal approximations.

\begin{figure}
\includegraphics[scale=0.5]{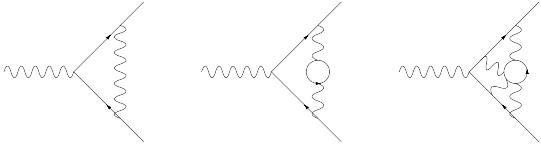}
\caption{Examples of connected subdiagrams $G_c$ for soft 
emissions between hard outgoing particles in abelian perturbation theory.}
\label{exp1}
\end{figure}

\begin{figure}
\begin{center}
\includegraphics[scale=0.5]{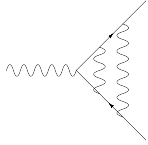}
\caption{Example of a disconnected subdiagram between two 
outgoing external lines, to be compared with the connected 
subdiagrams of Fig. \ref{exp1}.}
\label{discon}
\end{center}
\end{figure}

To show how this works, consider the path-integral representation 
of the free scalar Feynman propagator
\begin{equation}
  \label{eq:3}
  \Delta_F = \left[ {\rm i} (S - {\rm i} \varepsilon) \right]^{-1} \, , 
  \qquad S = (- \Box_x + m^2) \, ,
\end{equation}
which in momentum space reads
\begin{eqnarray}
  \tilde{\Delta}_F (p_f^2) & = &\frac12 \int_0^\infty d T \, 
  \frac{\matelem{p_f}{U(T)}{x_i}}{\bracket{p_f}{x_i}} 
  \nonumber \\ & = & - \frac{{\rm i}} {p_f^2 + m^2 
  - {\rm i}\varepsilon} \, .
\label{propmom}
\end{eqnarray}
In Eq.~(\ref{propmom}) we may introduce a path-integral 
representation for the matrix element
\begin{multline}
\label{eq:10}
  \matelem{p_f}{U(T)}{x_i} = {\rm e}^{- {\rm i} p_f
  x_i - {\rm i} \frac12 \left( p_f^2 + m^2 \right) T }
  \\ \times \, 
  \int_{x(0) = 0}^{p(T) = 0} \measure{p} \, \measure{x} 
  {\rm e}^{ {\rm i} \int_0^T d t \left( p \dot{x} - 
  \frac12 p^2 \right) } \, .
\end{multline}
The path integral is over all paths $x(t)$ with associated momentum
$p(t)$, starting at fixed position $x_i$ and ending with final 
momentum $p_f$. It is not hard to generalize this by including 
an abelian gauge field. One gets
\begin{multline}
  \label{eq:14}
  \matelem{p_f}{U(T)}{x_i} = \int_{x (0) = x_i}^{p(T) = p_f} 
  \hspace{-1mm} \measure{p} \, \measure{x} \, \exp 
  \hspace{-1mm} \Bigg[- {\rm i} \, p(T) x(T) \\ 
  + \, {\rm i} \int_0^T d t \left(p \dot{x} - \frac12 (p^2 + m^2)
  + p \cdot A \right. \\ 
  + \left. \frac{{\rm i}}{2} \partial \cdot A - \frac12 A^2
  \right) \Bigg] \, . 
\end{multline}
One may now express the $n$-point Green function correlating
$n$ hard particles by assuming a factorized form (which is known 
to be exact in the eikonal approximation), where hard fields
are implicitly integrated out, while the path integral over soft fields
yields eikonal and NE Feynman rules for soft emissions. We write
\begin{multline}
  G (p_1, \ldots, p_n) = \int \measure{A^{\mu}_s} \,
  H (x_1, \ldots, x_n) \\ \times
  \matelem{p_1}{(S - {\rm i} \varepsilon)^{-1}}{x_1} \ldots
  \matelem{p_n}{(S - {\rm i} \varepsilon)^{-1}}{x_n} \, ,
\label{eq:16}
\end{multline}
where $H$ collects all hard interactions, and there is an implicit 
integration over the coordinates $x_i$. Notice that the one-particle 
path integrals for each external line are functionals of the soft fields
which are then integrated explicitly in Eq.~(\ref{eq:16}). To extract 
a scattering amplitude, we must truncate the external lines. Each 
external line then carries a factor of the form
\begin{equation}
\label{eq:20}
  (p_f^2 + m^2) \matelem{p_f}{( S - {\rm i}
  \varepsilon)^{-1}}{x_i} \equiv {\rm e}^{ - {\rm i} p_f x_i} 
  f (\infty) \, ,
\end{equation}
where, after carrying out the $\measure{p}$ integration,
\begin{eqnarray}
  && \hspace{-5mm} f (\infty) = \int_{x(0) = 0} \hspace{-1mm}
  \measure{x} \, 
  \exp\Bigg[ {\rm i} \int_0^\infty d t
  \Bigg( \frac12 \, \dot{x}^2 \nonumber \\ && 
  + \, (p_f + \dot{x}) 
  \cdot  A \left(x_i + p_f t + x(t) \right) \nonumber \\ && +
  \frac{{\rm i}}{2} \partial \cdot A (x_i + p_f t + x) 
  \Bigg) \Bigg] \, . 
\label{eq:21}
\end{eqnarray}
From the point of view of the path integral over $A_s$ in
Eq.~(\ref{eq:16}) this is a collection of 1-point vertices for 
$A_s$, {\it i. e.} source terms. The scattering amplitude now 
reads
\begin{multline}
\label{eq:19}
  S(p_1, \ldots, p_n) = \int \measure{A^\mu_s} \,
  H(x_1, \ldots, x_n) \\ \times \,  
  {\rm e}^{- {\rm i} p_1 x_1} f_1(\infty) \ldots 
  {\rm e}^{- {\rm i} p_n x_n} f_n(\infty) \, \,
  {\rm e}^{ {\rm i} S[A_s]} \, .
\end{multline}
Each external line factor contains $A_s$ sources distributed along
paths which are summed over in the one-particle path integrals. 
The $A_s$ path integral then connects these sources with propagators
in all possible ways. The resulting gauge boson subdiagrams are either 
connected (as in Fig.~\ref{exp1}) or disconnected (as in 
Fig.~\ref{discon}).
At this point we can use the textbook result that, after performing
the $A_s$ path integral, the scattering amplitude is the exponent 
of connected graphs. This gives a simple and direct proof of 
exponentiation for eikonal diagrams, and for a class of NE 
contributions: those that obey the factorization assumed in
Eq.~(\ref{eq:16}).

The combinatorics underlying this statement assume commuting
sources. When considering a non-abelian gauge field this no longer 
holds, as each source carries a non-abelian charge. In order to use 
the same approach nevertheless, we employ a method that 
effectively shortcuts much of the combinatorial analysis for 
exponentiation: the so-called replica trick, which is used at times 
in statistical physics. Let us briefly sketch how the method works.

The Green functions of a given quantum field theory are described 
by the generating functional
\begin{equation}
  Z [J] = \int \measure{\phi} \, \, {\rm e}^{ {\rm i} S[\phi]
  + {\rm i} \int J \phi} \, ,
\label{Z}
\end{equation}
where $J$ is a source for the field $\phi$, and $S$ is the classical 
action. Now consider defining $N$ replicas of the theory,
involving fields $\phi_i$ ($i \in \{1, \ldots, N \}$). The
generating functional becomes
\begin{multline}
  Z_N [J] = \int \measure{\phi_1} \ldots \measure{\phi_N} \\
  \times {\rm e}^{ {\rm i} S[\phi_1] + {\rm i} \int J \phi_1}
  \ldots {\rm e}^{ {\rm i} S[\phi_N] + {\rm i} \int J \phi_N} \,,
\label{ZN}
\end{multline}
which clearly satisfies
\begin{equation}
  Z_N [J] = \left( Z[J] \right)^N \, .
\label{ZN2}
\end{equation}
The Feynman rules for each field are identical, and there are no 
interactions between the different replicas of the fields: thus, there 
can be no more than one field in each connected Feynman diagram, 
and connected diagrams come in $N$ copies. By the same reasoning,
disconnected diagrams containing $k \geq 2$ connected components 
come in $N^k$ copies. It follows that the sum of all connected 
diagrams is proportional to the number of replicas, $\sum G_c
\propto N$. Furthermore, no disconnected diagrams contribute 
terms proportional to $N$. From Eq.~(\ref{ZN2})
one has
\begin{equation}
  Z_N [J] = 1 + N \log(Z[J]) + {\cal O}(N^2) \, ,
\label{ZN3}
\end{equation}
which leads us to conclude that
\begin{equation}
  \sum G_c = \log \left( Z[J] \right) \, .
\label{Gcsum2}
\end{equation}
Finally, we may write this as
\begin{equation}
  Z [J] = \exp \left[ \sum G_c \right] \, ,
\label{Z2}
\end{equation}
and set $N = 1$. For non-abelian gauge theory we note that 
although the fields are replicated, the gauge group is not, so that 
all replicas live in the same Lie algebra. 

Let us consider the case of two external colored particles,
corresponding to color-singlet process such as Drell-Yan. The 
scattering amplitude reads
\begin{multline}
  S(p_1, p_2) = H (p_1,p_2) \int \measure{A^\mu_s} \, \\
  \times f_1^{i j_1} (\infty) \, f_2^{i j_2} (\infty) 
  \, {\rm e}^{ {\rm i} S[A_s]} \, ,
\end{multline}
where the external lines factors are path-ordered
exponentials similar to Eq.~(\ref{eq:21}).   We can now use 
the combinatorial power of the path integral by replicating the 
gluon field, and defining a replica-ordering operator ${\cal R}$ 
such that
\begin{multline}
\label{eq:2}
  \prod_{i = 1}^N {\cal P} \exp \left[ \int d x \cdot A_i (x) \right]
  \\ = {\cal R}{\cal P} \, \exp \left[ \sum_{i = 1}^N \int d x 
  \cdot A_i (x)\right] \, ,
\end{multline}
where
\begin{equation}
  {\cal R} \left[A_i (x) A_j (y) \right] = \left\{
  \begin{array}{c} A_i (x) A_j (y) \, , \quad i \leq j
  \\ A_j (y) A_i (x) \, , \quad i>j \end{array} \right. \, .
\label{Rdef}
\end{equation}

The classical path $x(t)$ in Eq.~(\ref{eq:21}) is a straight line, 
leading to the standard eikonal approximation. By combining the 
two half-infinite paths into one path, we can categorize NE 
corrections using 1-dimensional field theory on the path. 
Two-point correlators correspond to path fluctuations, and 
lead precisely to NE Feynman rules for the soft gauge field in 
4-dimensional Minkowski space. 

The replica trick allows us to decide which diagrams
connecting the sources on the path occur in the exponent,
and, moreover, what their color factors are. In the eikonal
approximation we recover precisely Gatheral's webs
\cite{Sterman:1981jc,Gatheral:1983cz,Frenkel:1984pz}, 
with a non-recursive recipe to determine their modified color 
factors. Moreover, we identify at NE accuracy which new diagrams 
occur in the exponent (NE webs), and what their color factors are. 
Note however that not all NE corrections exponentiate: emissions 
that connect the hard function $H$  to the external lines, violating 
the factorization assumed in Eq.~(\ref{eq:16}), and which are 
associated with the Low-Burnett-Kroll theorem \cite{Low:1958sn,Burnett:1967km,DelDuca:1990gz}, do not exponentiate. 
They can, however, be organized into an iterative pattern, as shown 
in \cite{Laenen:2008gt}.

\section{Diagrammatic approach}
\label{sec:diagr-appr}

For a different perspective on these results, one can follow a purely
diagrammatic approach. For abelian gauge theory, in the eikonal
approximation, one can consider all diagrams with an arbitrary
number of photon exchanges (possibly via closed fermion loops).
Such diagrams will in general contain disconnected pieces. Upon
expanding all propagators and vertices to leading power in the soft 
momenta, and summing over all permutations  $\pi$ of the emitted
photons, one can decorrelate all photon emissions from each other 
by using the eikonal identity
\begin{multline}
  \sum_\pi \frac{1}{p \cdot k_{\pi_1}} \, \frac{1}{p \cdot
  (k_{\pi_1} + k_{\pi_2})} \, \ldots \\
  \times \frac{1}{p \cdot(k_{\pi_1} + k_{\pi_2} + \ldots 
  k_{\pi_n})} = \prod_i \frac{1}{p \cdot k_i} \, .
\label{eikonalid}
\end{multline}
Simple combinatoric arguments then show that the full amplitude 
${\cal A}$, dressing the radiationless amplitude ${\cal A}_0$
with multiple soft photon radiation, takes the form
\begin{multline}
  {\cal A}  =  {\cal A}_0 \sum_{\{N_i\}} \prod_i 
  \frac{1}{N_i!} \, [G_c^{(i)}]^{N_i} \, ,
\label{amp4}
\end{multline}
where $G^{(i)}_c$ is a connected photon subdiagram joining
the external lines, and $N_i$ is the multiplicity of this subdiagram 
occurring in the full diagram. Eq.~(\ref{amp4}) clearly displays an exponentation of the form of Eq.~(\ref{Z2}).

For non-abelian gauge theories one must use a generalization of 
the eikonal identity
\begin{eqnarray}
  && \sum_{\tilde{\pi}} \frac{1}{p \cdot k_{\tilde{\pi}_1}} \,
  \frac{1}{p \cdot (k_{\tilde{\pi}_1} + k_{\tilde{\pi}_2})}
  \ldots \nonumber \\ && \qquad \quad \times \frac{1}{p 
  \cdot(k_{\tilde{\pi}_1} + \ldots + k_{\tilde{\pi}_n})} \\
  && \quad = \, \, \prod_g \left[ \frac{1}{p \cdot k_{g_1}} \, 
  \frac{1}{p \cdot (k_{g_1} + k_{g_2})} \right. \notag 
  \nonumber \\
  && \qquad \quad \left. \times \ldots \frac{1}{p \cdot(k_{g_1} 
  + \ldots + k_{g_m})} \right] \, , \nonumber
\label{eikid2}
\end{eqnarray}
where the product is over ``groups'' $g$, defined as projections 
of webs onto each eikonal line. The permutations $\tilde{\pi}$ 
are restricted to keep the ordering of gluon attachments within 
each group fixed. This product actually gives the set of eikonal 
gluon amplitudes the structure of a shuffle algebra~\cite{Weinzierl:2010ps}. Using this algebraic structure one can set up an 
inductive proof of exponentiation, both for eikonal and NE 
approximations \cite{appear1}, which confirms and complements
the discussion of~\cite{Laenen:2008gt}. 

The NE approximation, for those terms that exponentiate according
to our discussion, can be implemented in practice through a set of 
effective Feynman rules. As a test, we used these rules to compute 
the abelian terms of the double-real emission contribution to the 
Drell-Yan process at two loops, we combined the result with the 
corresponding phase space and compared with the exact 
result \cite{DYexact}. We find
\begin{eqnarray}
  && \hspace{-5mm} K^{(2)}_{{\rm NE}} (z) \, = \, \left( 
  \frac{\alpha_s C_F}{4 \pi} \right)^2
  \Bigg[ \frac{1024 \, {\cal D}_3 (z)}{3} \nonumber \\
  && \hspace{-3mm} - \, \frac{1024 \log^3(1- z)}{3} 
  + 640 \log^2(1- z) \nonumber \\
  && \hspace{-3mm} + \, \frac{512 \, {\cal D}_2 (z) - 
  512 \log^2(1- z) + 640 \log(1- z)}{\epsilon} \nonumber \\
  && \hspace{-3mm} + \, \frac{512 \, {\cal D}_1(z) - 
  512 \log(1 - z)}{\epsilon^2}  \label{NEtotal} \\
  && \hspace{-3mm} + \, \frac{256 \, {\cal D}_0(z) - 
  256}{\epsilon^3} \Bigg]  \, . \nonumber 
\end{eqnarray}
where ${\cal D}_k (z)$ are plus distributions, ${\cal D}_k (z) = 
[\ln^k (1 - z)/(1 - z)]_+$. The result is in complete agreement 
with \cite{DYexact}.

\section{Conclusions}
\label{sec:conclusions}

How to resum sub-leading logarithms at next-to-eikonal accuracy is
an unsolved problem, relevant for phenomenological applications 
\cite{soar,kls}. An approach based on path integrals and using the 
replica trick yields much insight into exponentiation of these 
contributions in terms of (next-to-)eikonal webs, and it is 
corroborated by a diagrammatic analysis. We hope in future 
to extend our study to the level of full cross sections, and 
to processes involving more hard colored partons.

\end{document}